\documentclass[twocolumn,showpacs,preprintnumbers,amsmath,amssymb,latexsym,prd]{revtex4-1}
\usepackage[utf8]{inputenc}
\usepackage{graphicx}
\usepackage{color}
\usepackage{amsmath}
\usepackage[left,running]{lineno}
\usepackage[utf8]{inputenc}
\usepackage{float}

\newcommand{\f}[2]{\frac{#1}{#2}}

\renewcommand{\Re}{{\rm Re}\,}

\newcommand{\tr}{{\rm tr}\,}

\begin{document}

\title{High statistics lattice study of stress tensor correlators in pure $SU(3)$ gauge theory}

\author{
Sz.   Bors\'anyi$^a$,
Z.    Fodor$^{abc}$,
M. Giordano,$^{c}$,
S. D. Katz$^{c}$,
A. P\'asztor$^{a}$,
C. Ratti$^{d}$,
A. Sch\"afer$^{e}$,
K. K. Szab\'o$^{ab}$,
B. C. T\'oth$^{a}$}
\address{
$^{a}$University of Wuppertal, Department of Physics, Wuppertal D-42097, Germany\\
$^{b}$J\"ulich Supercomputing Centre, J\"ulich D-52425, Germany\\
$^{c}$E\"otv\"os University, Budapest 1117, Hungary\\
$^{d}$ Department of Physics, University of Houston, Houston, TX 77204, USA \\
$^{e}$University of Regensburg, Regensburg D-93053, Germany
}

\begin{abstract}
We compute the Euclidean correlators of the stress tensor in pure $SU(3)$
Yang-Mills theory at finite temperature at zero and finite spatial momenta with
lattice simulations. We perform continuum extrapolations using
$N_\tau=10,12,16,20$ lattices with renormalized anisotropy 2. 
We use these correlators to estimate the shear viscosity of 
the gluon plasma in the deconfined phase. For $T=1.5T_c$ we obtain
$\eta/s=0.17(2)$.
\end{abstract}

\maketitle

\section{Introduction}

Since relativistic hydrodynamics is quite successful in the interpretation of
heavy ion experiments~\cite{Teaney:2001av, Romatschke:2007mq,
Romatschke:2009im, Schafer:2009dj, Heinz:2013th} it would be of great interest
to calculate the shear viscosity of the quark gluon plasma from first
principles.

In classical transport theory, the shear viscosity to entropy density ratio for a dilute gas at temperature T is  $\eta/s \sim T l_{\rm{mfp}} \bar{v}
\sim \frac{T \bar{v}}{n \sigma}$, where $l_{\rm{mfp}}$ is the mean free path,
$\bar{v}$ is the mean speed, $n$ is the particle number density and $\sigma$
the cross section. For a weakly interacting system $\sigma$ is small, and $\eta/s$ is
expected to be large. In particular, for a free gas, $\eta/s$ is infinite. On
the other hand, for a strongly interacting system $\eta/s$ is expected to be small
~\cite{Kovtun:2004de}. As heavy ion phenomenology points to 
a rather small viscosity~\cite{Teaney:2001av,
Romatschke:2007mq, Romatschke:2009im, Schafer:2009dj, Heinz:2013th}  
a non-perturbative calculation of the shear viscosity would be a
great success. 

One possible route to determine the viscosity is
through the Kubo formula, relating transport coefficients to the zero-frequency
behavior of spectral functions.  The relevant Kubo formula for the shear
viscosity is: 
\begin{align} \label{eq:Kubo} \eta(T) = \pi \lim_{\omega \to 0}
\lim_{\mathbf{k} \to 0} \frac{\rho_{ijij}(\omega,\mathbf{k},T)}{\omega} \, 
\end{align} 
where $\rho_{ijij}(\omega,\mathbf{k},T)$ is the spectral function corresponding to
the energy momentum tensor at the specified spatial indices $i\ne j$.  The
direction of the momentum is $j$.  In this paper, we will assume without any
loss of generality, that the external momentum is in the $3$rd direction, while
the zeroth direction is the (Euclidean) time.  By choosing a matching $i$ index
we will consider the component $\rho_{1313}(\omega,\mathbf{k},T)$.

  In general, the correlator of the energy momentum tensor $T_{\mu\nu}$ is
given in Euclidean space-time as
\begin{equation}
  C_{\mu\nu,\rho\sigma}(\tau,\vec x) = \int \langle T_{\mu\nu}(\tau',\vec {x'})
T_{\rho\sigma}(\tau'+\tau,\vec {x'}+\vec{x}) \rangle d\tau' d\vec {x'}\,,
\end{equation}
which is a direct observable on the lattice.  Its Fourier transform is related
to the spectral function by an integral transform

\begin{align}
\label{eq:sumrule}
C_{\mu \nu, \rho \sigma}(\tau, \mathbf{q}) = 
\int_{0}^{\infty} d \omega \rho_{\mu \nu, \rho \sigma}(\omega, \mathbf{q}, T) 
K( \omega, \tau ; T) \rm{,}
\end{align}
with the kernel
\begin{equation}
    K(\omega, \tau; T) = \frac{\cosh \left( \omega \left( \tau - 1/(2T) \right)   \right) }{ \sinh \left( \omega / (2T) \right)} \rm{.}
\end{equation}

Both early ~\cite{Karsch:1986cq,Nakamura:2004sy,Meyer:2007ic} and more
recent~\cite{Astrakhantsev:2017uld} lattice studies of the viscosity
used the Kubo formula (\ref{eq:Kubo}). In this approach the
integral transform (\ref{eq:sumrule}) has to be inverted.
For $T \gg \omega$ the kernel behaves like $e ^{-\omega \tau}$, i.e. our task
is similar to inverting a Laplace transform numerically. It is well known that
such an approach is bound to face great difficulties. There are two
interrelated problems:
\begin{enumerate}
\item Equation (\ref{eq:sumrule}) is a Fredholm equation of the first 
       kind, which for most kernels very ill-posed. The difficulty can 
       intuitively be compared to the process of de-blurring an image. 
       Also, in particular, both the Laplace kernel, and our kernel 
$K(\omega, \tau; T)$ are known to lead to a very ill-conditioned
inverse problem~\cite{Meyer:2011gj}. 
\item For the particular case of the viscosity, the signal in the stress-energy
tensor is strongly dominated by the high frequency part of the spectral
function~\cite{Aarts:2002cc}. This makes reconstruction even harder 
as the blurring character of the integral transform (point 1) mixes 
the contributions from the high and low
$\omega$ part of the spectral function in the measured Euclidean correlator.
\end{enumerate}
To see how bad a particular inversion
problem is, it is very instructive to look at the spectral function for the free theory, as this will
correspond to the asymptotic behavior of the spectral function in the continuum theory, because of
asymptotic freedom. To get the asymptotic behavior up to a constant, one only has to perform simple
dimensional analysis. For $\rho_{1313}$ this leads to an asymptotic $\omega^4$ behavior, making the 
UV contamination especially severe. 

\begin{figure}[t!]
\begin{center}
\includegraphics[trim={1.8cm 0 0 0},clip, width=0.52\textwidth]{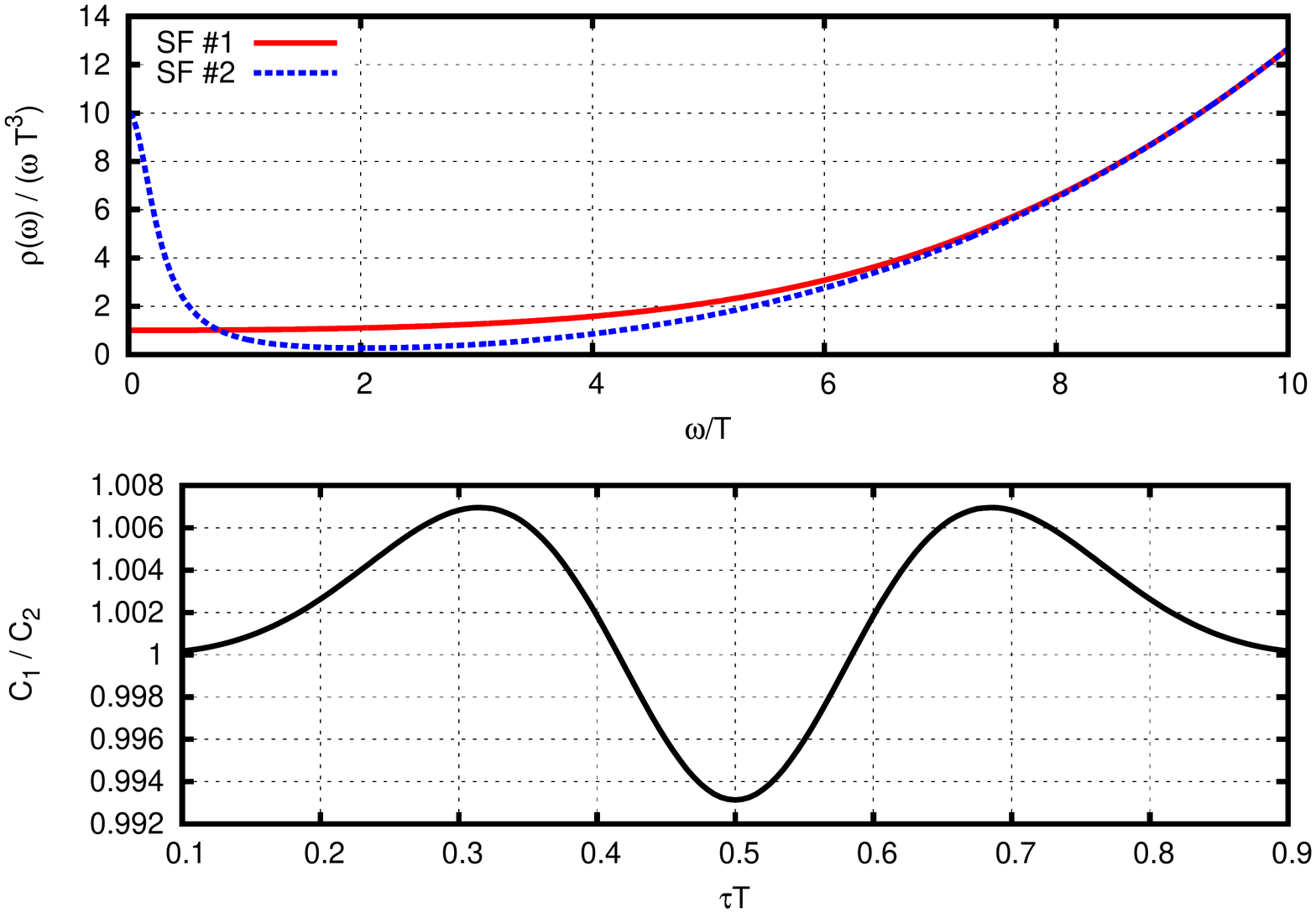}
\caption{\label{fig:scary1} The Euclidean correlator corresponding to the spectral function appearing in equation (1) 
is very insensitive to its IR features. To illustrate this, we show two different 
spectral functions, with the same UV, but different 
IR features (top) and the ratio of the corresponding Euclidean correlators (bottom). 
The viscosities are different by a factor of $10$, but the Euclidean 
correlators differ by less than $1\%$.}  
\end{center}
\end{figure}

To see an honest illustration of these problems for $\rho_{1313} \sim
\omega^4$, look at Figure \ref{fig:scary1}, where we illustrate how insensitive
the Euclidean correlator is to the IR features of the spectral function. There
we show two different spectral functions, with a factor of $10$ difference in
the viscosity, that nevertheless lead to sub percent differences in the
corresponding Euclidean correlators. The two mock spectral functions in Figure
\ref{fig:scary1} are actually both physically motivated. The featureless
spectral function (\#1 in Fig.~\ref{fig:scary1}) is reminiscent of the one
obtained from calculations in $\mathcal{N}=4$ SYM theory, with AdS/CFT
methods~\cite{Teaney:2006nc}, while the spectral function exhibiting a
Lorentzian peak at $\omega=0$ is reminiscent of the kind of results one obtains
from leading log kinetic theory calculations in QCD itself ~\cite{Hong:2010at}.
Since the AdS/CFT calculation is a strong coupling calculation in the wrong
theory, while the kinetic theory calculation is a calculation in the wrong
regime of QCD, we do not know a priori which type of spectral function we can
expect for QCD in the phenomenologically relevant temperature range, so a fully
controlled calculation of the viscosity from the Kubo formula would necessarily
need to distinguish between these two scenarios. 

Note that for 
Figure~\ref{fig:scary1} we assumed that the asymptotic behavior of the spectral function
is known completely accurately and there are no features of the spectral function at intermediate
frequencies (e.g. no glueballs or remnants of melted glueballs). 
Though there has been progress in perturbative calculations of the UV part of the spectral 
functions~\cite{Schroder:2011ht, Zhu:2012be, Vuorinen:2015wla}, 
these assumptions are optimistic. 
Still, the fact that under these assumptions 
an order of magnitude difference in the viscosity leads to 
less than 1$\%$ difference in the Euclidean correlators nicely illustrates our point. 

The bottom line of this discussion is 
that, for a credible lattice estimate of the viscosity,
a high level of precision is
necessary for the Euclidean correlators,
especially if we want to use equation
(\ref{eq:Kubo}), like it was done in Refs.
~\cite{Karsch:1986cq,Nakamura:2004sy,Meyer:2007ic, Astrakhantsev:2017uld}.

The situation is much better for the correlators of conserved charges, 
appearing in the electric conductivity~\cite{Aarts:2007wj, Ding:2010ga,
Aarts:2014nba, Ding:2016hua} and heavy quark
diffusion~\cite{Petreczky:2005nh,Jakovac:2006sf,Borsanyi:2014vka,Francis:2015daa} calculations.
In those cases the spectral function at large $\omega$ only grows like
$\omega^2$, making the UV contamination problem less severe.  It was an
important realization of~Refs.~\cite{Meyer:2008gt, Meyer:2011gj} that even for
the case of the shear viscosity, the asymptotic $\omega^4$ behavior can be made
better, only $\omega^2$, by utilizing the following Ward identity:
\begin{align}
\label{eq:Ward}
-\omega^2 \rho_{0101} = \mathbf{q}^2 \rho_{1313} \rm{,}
\end{align} 
and using the $\rho_{0101}$ correlator, instead of the $\rho_{1313}$. This
would make the asymptotic behavior of the shear viscosity spectral function
only as bad as that of the electric conductivity. But there are crucial
differences as well. In the continuum we 
have the thermodynamic identity \footnote{See Appendix A} $\left< T_{01}T_{01} \right> (\tau,\mathbf{q}=0)/T^5=s/T^3$.
This means that we need nonzero momenta to obtain information about the viscosity from this correlator.

Even with this knowledge, the calculation of the viscosity is still much more difficult than that 
of the electric conductivity. 
The source of the difficulty is the fact that the
stress-energy tensor correlators $C(\tau)$ have a quickly degrading signal as
$\tau$ is increased beyond a few lattice spacings. Usually, the width of the
distribution for these observables in a Monte Carlo simulation is much 
larger than the value, at least near the 
middle point $\tau T = 1/2$, the very point 
where the correlator has its highest sensitivity to transport.
Thus, the physically most relevant quantity is evaluated as an average 
of wildly fluctuating contributions (with fluctuating sign), which is
typically the characteristic of a sign problem.

For the quenched case this problem can be ameliorated 
by using the multilevel algorithm~\cite{Luscher:2001up, Meyer:2002cd}. 
This algorithm depends crucially on the locality of the 
action, and therefore it proved to be hard to generalize for 
dynamical fermions. Some progress in this regard has been 
made recently in \cite{Ce:2016idq, Ce:2016ajy}. 
Nevertheless, at least in the quenched case, high statistical 
precision can be achieved via the multilevel algorithm.

The study of cut off effects of these correlators is rather limited in the
literature. The tree level improvement coefficients for the plaquette action
and two different discretizations of $T_{\mu \nu}$ where calculated
in~\cite{Meyer:2009vj}.  So far no calculations of these correlators are
available with three lattice spacings in the scaling regime.

In this paper, we take steps towards achieving the high precision necessary for the calculation
of the shear viscosity, by inverstigating several technical aspects of such a calculation. Namely:
\begin{itemize}
\item Utilizing a different gauge action, the tree level Symanzik-improved action, as opposed to the plaquette action used in previous studies.
\item Studing the continuum limit behavior by simulating at different values of the lattice spacing $N_t=10,12,16$ and $20$.
\item Calculating the $w_0$ scale with high precision.
\item Using the Wilson flow for anisotropy tuning, as advertised in~\cite{Borsanyi:2012zr}
\item Using shifted boundary conditions for the renormalization of the energy momentum tensor, a technique that was worked out for the 
isotropic case in~\cite{Giusti:2015daa}. Here, we utilize it for an anisotropic lattice.
\item Calculating the tree level improvement coefficients for the Symanzik-improved gauge action.
\end{itemize}

Throughout this paper, we will mostly focus on the calculation of the energy-momentum tensor, and not the inversion method
for reconstructing the spectral function. We believe this to be an important first step. Before the inversion can be done,
one needs to have reliable results for the correlator itself. Nevertheless, in the end we give an estimate of the viscosity,
using a similar hydrodynamics motivated fit ansatz as some previous studies~\cite{Meyer:2011gj}.

\section{Lattice calculation of the correlators}

Our calculation uses the tree-level Symanzik-improved gauge-action:
\begin{equation}
\begin{aligned}
S_{\rm tli} &= \beta\sum_n
\sum_{\mu<\nu}\f{\lambda_\mu\lambda_\nu}{\lambda_{\bar\mu}\lambda_{\bar\nu}}
\Big[1- \f{1}{N_c}\Re\tr {\cal U}_{\mu\nu}(n)\Big]\,,\\
{\cal U}_{\mu\nu}(n) &= c_0 W_{\mu\nu}(n;1,1) \\ & \phantom{=}+ c_1W_{\mu\nu}(n;2,1) 
+ c_1W_{\mu\nu}(n;1,2)\,, 
\end{aligned}
\end{equation}
where $\beta =\f{2N_c}{g^2}$. $\bar\mu$ and $\bar\nu$ are the complementer
indices for $\mu$ and $\nu$, such that $\bar\mu < \bar\nu$ and the
four indices $\mu,\nu,\bar\mu,\bar\nu$ are a permutation of 0,1,2,3.
Here $W_{\mu\nu}(n;a,b)$ are Wilson loops around rectangular $a\times b$ paths.
Finally, $c_0=\f{5}{3}$ and $c_1=-\f{1}{12}$. 

The anisotropy parameters are $\lambda_1=\lambda_2=\lambda_3=1$, $\lambda_4=\xi_0$, with $\xi_0$ the bare anisotropy.
For our study we use anisotropic lattices with renormalized anisotropy $\xi_R = 2$. For anisotropy tuning we
use the Wilson flow technique introduced in \cite{Borsanyi:2012zr}. The procedure for anisotropy tuning will
be detailed later.

We use a multilevel algorithm (more precisely, a two-level algorithm
\cite{Meyer:2003hy}) to reduce errors near $\tau T=0.5$.

We use the clover discretization
of the energy momentum tensor, mainly because the center of the 
operator is always located on a site, 
therefore the separation of the operators is always an integer 
in lattice units. If one were to use the plaquette discretization
there would be a component that is defined for integer separations
and one that is defined for half integer separations, and one would 
need an interpolation to add them together. This would lead to the
appearance of a systematic error coming from the interpolation, that
we want to avoid.

Following the line of previous studies we use the two-level algorithm,
but now with a tree-level Symanzik improvement. \footnote{Note, 
that this 
combination was already used in the literature for the 
calculation of static quark potentials \cite{Mykkanen:2012dv}.}
Thus we have thick
layers (having a width of a full temporal lattice spacing) between the
blocks in the inner update.

We have ensembles at two different temperatures: $1.5T_c$ and $2T_c$, and the
following lattice geometries: $N_z\times N_y^2 \times N_t=$
 $80 \times 20^2 \times 20$, $64 \times 16^2
\times 16$, $48 \times 12^2 \times 12$, $40 \times 10^2 \times 10$.
The long spatial direction is needed so that we can have small spatial momenta, 
to justify our hydrodynamics motivated fit ansatz below.

\subsection{Statistics}

As was explained in the introduction, the correlators are needed to a very high
precision, if one wants to have useful information on transport. In the pure
$SU(3)$ theory this can be achieved by using a multilevel
algorithm~\cite{Luscher:2001up} and high statistics.  Earlier lattice studies of
the viscosity~\cite{Karsch:1986cq,Nakamura:2004sy,Meyer:2007ic,
Astrakhantsev:2017uld} also use a multilevel algorithm. 

The main difference is -- apart from the higher statistical precision --
that we are working with four different lattice spacings, which
allows us to study the correlators in the continuum limit. 
Our statistics is summarized in Table~\ref{table:stat}.

\begin{table}
\begin{center}
\begin{tabular}{  c | c | c | c | c  }
 & $40 \times 12^2 \times 10$ & $48 \times 12^2 \times 12$ & $64 \times 16^2 \times 16$ & $80 \times 20^2 \times 20$ \\
\hline
  $1.5 T_c$ & 2.03M    & 4.99M    & 5.11M    & 1.63M    \\
  $2.0 T_c$ & 2.07M    & 4.86M    & 6.31M    & 1.57M    \\
\end{tabular}
\caption{\label{table:stat}Number of measurements (millions) of the
energy-momentum tensor correlators at the simulation points. Between every
measurement there are 100 regular updates and 500 inner multilevel updates.}
\end{center}
\end{table}

\begin{figure}[h!]
\begin{center}
\includegraphics[width=0.5\textwidth]{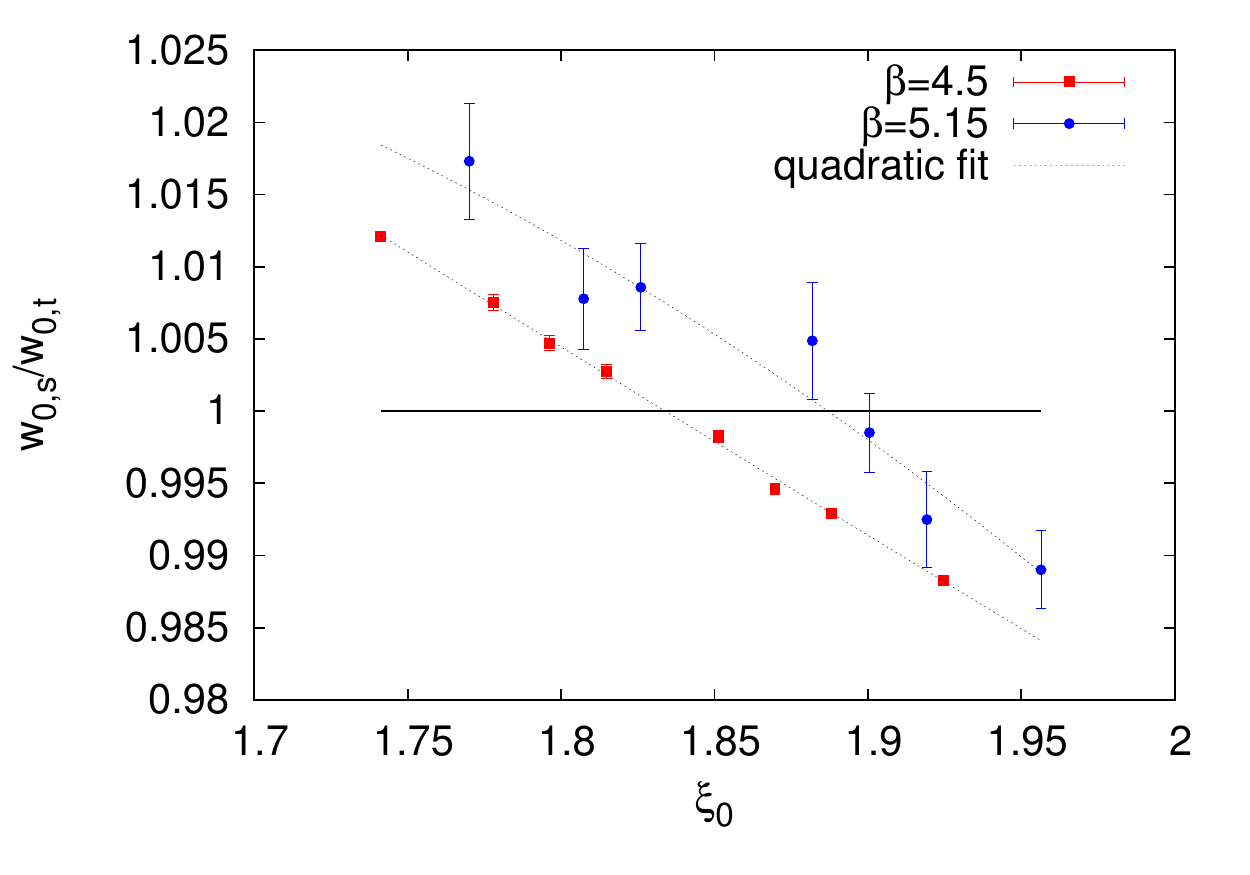} \\
\includegraphics[width=0.5\textwidth]{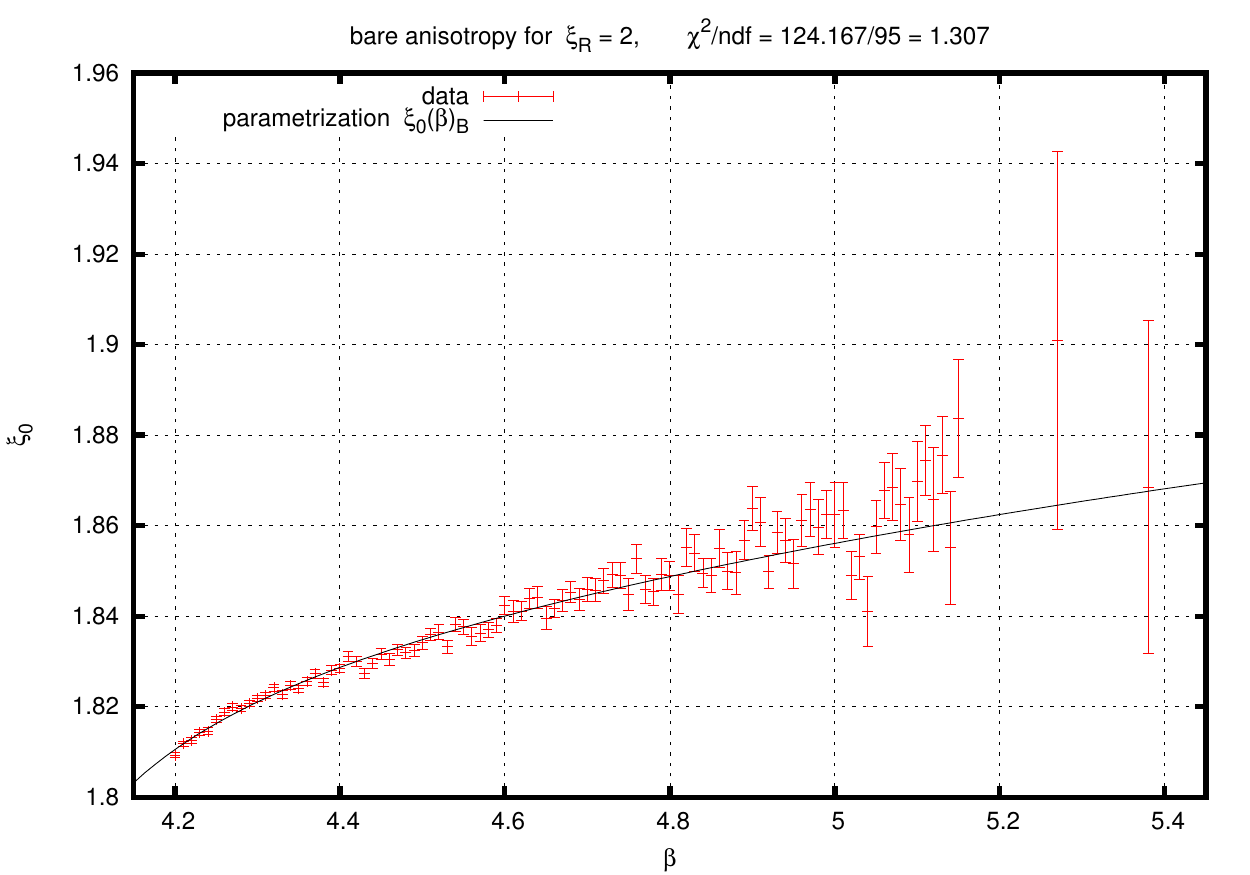}
\caption{
\label{fig:xi}
Top: Anisotropy tuning with simulations at different bare anisotropies. The tuned bare anisotropy corresponds to $w_{0,s}/w_{0,t}=1$.
Bottom: Parametrization of the bare anisotropy used for our simulations.}  
\end{center}
\end{figure}

\subsection{Anisotropy tuning and scale setting}
To fix the anisotropy we use the method introduced in~\cite{Borsanyi:2012zr}. 
The bare anisotropy $\xi_0(\beta)$ is tuned so that $\xi_R\equiv 2$. For the tuning we define 
a \textit{spatial} and a \textit{temporal} $w_0$ scale:
\begin{align}
\left[\tau\frac{d}{d\tau} \tau^2 \langle E_{ss}(\tau) \rangle\right]_{\tau=w_{0,s}^2} = 0.15\,,\\
\left[\tau\frac{d}{d\tau} \tau^2 \langle E_{ts}(\tau) \rangle\right]_{\tau=w_{0,t}^2} = 0.15\,,
\end{align}
with
\begin{align}
E_{ss}(\tau) = \frac{1}{4}\sum_{x,i\ne j} F_{ij}^2(x,\tau)\,,\\
E_{st}(\tau) = \xi_R^2\frac{1}{2}\sum_{x,i}F_{i4}^2(x,\tau)\,.
\end{align}
To tune the anisotropy we use the following procedure:
\begin{enumerate}
 \item We simulate the $SU(3)$ theory at fixed $\beta$ and several bare anisotropies around our estimate~\cite{Borsanyi:2012zr},
 targeting $\xi_R=2$
 \item We calculate the gradient flow using $\xi_R=2$ and monitor $w_{0,x}/w_{0,t}$ as a function of $\xi_0$ (see Fig.~\ref{fig:xi}).
The correct tuning of the anisotropy is achieved when $w_{0,x}/w_{0,t}=1$.
 \item The $\xi_0(\beta)$ data set is fitted with a Padé formula. The fitted curve is plotted also in Fig.~\ref{fig:xi}. Our parametrization
reads:
\begin{equation}
    \xi_0(\beta) = 2.0 \left( 1 + \frac{6}{\beta} \frac{-0.0578007+0.2255046/\beta}{1.0-3.94044/\beta} \right) \rm{.}
\end{equation}
\end{enumerate}
We likewise fit $w_0(\beta)$, with a parametrization that interpolates smoothly between the two-loop running of the coupling and the 
lattice data:
\begin{align}
w_0(\beta)&= \exp \big[ -\frac{b_1}{2b_0^2} \log \left( \frac{\beta}{2 N_c b_0} \right) + \frac{\beta}{4 N_c b_0} \\ \nonumber
&- 3.51307817908059 \\ \nonumber
&- \frac{1}{-8.0963941698416 + 2.36701001378353 \beta} \big] \rm{,}
\end{align}
with $b_0=11N_c/48\pi^2$, $b_1=34N_c^2/768\pi^4$ and $N_c=3$. So far, we expressed the scale using $w_0$.
In order to be able to translate to $T_c$ scale we have to determine the combination $w_0 T_c$ in the continuum
limit. We did this using four different (isotropic) actions (Wilson, tree-level Symanzik, Iwasaki and DBW2).
With the exception of the last one, we found similar 
results using lattices up
to $N_\tau=12$ and a continuum limit using an $N_\tau^2$ as well as an
$N_\tau^4$ term. The uncontrolled systematics of the DBW2 result
is no surprise, this action is known to poorly sample topological
sectors, see for example \cite{DeGrand:2002vu}.

\begin{figure}[h!]
\begin{center}
\includegraphics[width=0.52\textwidth]{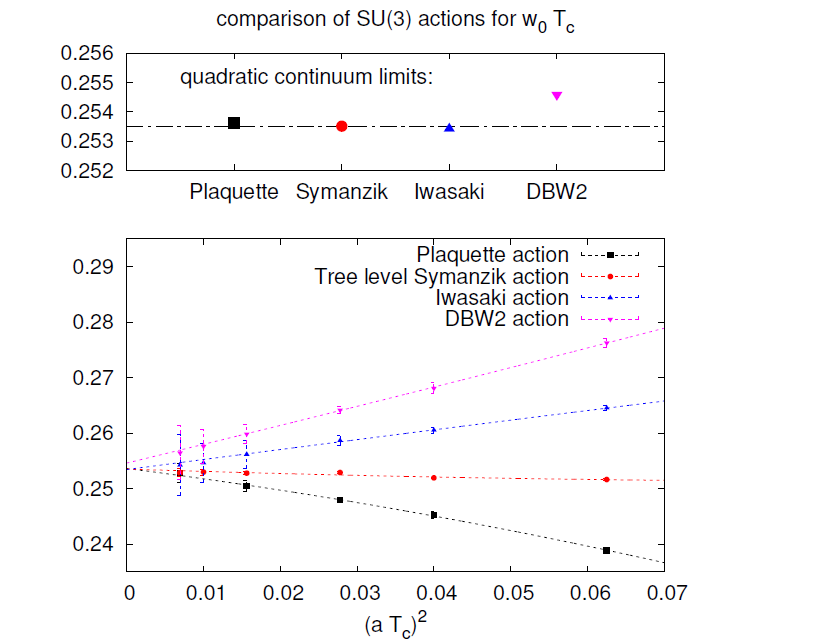}
\caption{
\label{fig:w0Tc}
Determination of $w_0 T_c$ from four different pure $SU(3)$ actions.}  
\end{center}
\end{figure}

In all cases $T_c$ was defined by the peak of the Polyakov
loop susceptibility. The summary plot for this study is shown in
Fig.~\ref{fig:w0Tc} For each action, the resulting jackknife error was very
small. Therefore we use the spread between the Wilson, Symanzik and Iwasaki
results as an error estimate instead, and use the Symanzik result (which lies central
between the others) as mean. We conclude that $w_0 T_c = 0.2535(2)$.

\subsection{Renormalization}

The translational symmetry is broken on the lattice. 
As a result, renormalization
factors appear between the lattice definition of the energy
momentum tensor $T_{\mu\nu}$ and the physical quantity. This factor
depends on the action, the discretization scheme in the
$T_{\mu\nu}$ observable and the lattice spacing (or the $\beta$ parameter).
Moreover, these factors are not the same for each component, since
the off-diagonal (sextet), the diagonal (triplet), and the trace (singlet)
correspond to different representations of the four-dimensional
rotation group. On an isotropic lattice one has three factors:

\begin{equation}
T_{\mu\nu}^R = Z_6 T^{[6]}_{\mu\nu} +  Z_3 T^{[3]}_{\mu\nu} + Z_1(T^{[1]}_{\mu\nu}-T^{[1]}_{\mu\nu}(T=0))
\end{equation}
where:
\begin{eqnarray}
T^{[6]}_{\mu\nu}& =&  \frac{1}{g_0^2} \sum_\sigma F^a_{\mu\sigma}F^a_{\nu\sigma} \rm{,}\\
T^{[3]}_{\mu\nu}& =& \delta_{\mu\nu} \frac{1}{g_0^2} 
\left\{ \sum_\rho F^a_{\mu\rho}F^a_{\nu\rho}- \frac{1}{4}\sum_{\rho,\sigma} F^a_{\rho\sigma}F^a_{\rho\sigma}\right\} \rm{,}\\
T^{[1]}_{\mu\nu}& =& \delta_{\mu\nu} \frac{1}{g_0^2} \sum_{\sigma,\rho} F^a_{\rho\sigma}F^a_{\rho\sigma} \rm{,}
\end{eqnarray}
and there is no summation over $\mu$ and $\nu$ in the above formulas.

We use the clover discretization of $F_{\mu\nu}^a$ and define our correlators
from the sextet (off-diagonal) components. In the presence of anisotropy, the
renormalization constant $Z_6$ splits into three different renormalization
constants:
\begin{eqnarray}
T_{01} &=&
\frac{Z_6^{ts}}{g_0^2} {F^a_{02}F^a_{12}} 
+\frac{Z_6^{ts}}{g_0^2} {F^a_{03}F^a_{13}} \rm{,}\\
T_{13} &=&
\frac{Z_6^{tt}}{g_0^2} {F^a_{01}F^a_{03}}
+\frac{Z_6^{ss}}{g_0^2} {F^a_{12}F^a_{32}} \rm{.}
\end{eqnarray}
In our renormalization procedure, we get $Z_6^{ts}$ from the thermodynamic
identity (\ref{eq:srenorm}), and we get the ratios $Z_6^{ss}/Z_6^{ts}$ and
$Z_6^{tt}/Z_6^{ts}$ from shifted boundary conditions.

\begin{figure}[h!]
\begin{center}
\includegraphics[width=0.5\textwidth]{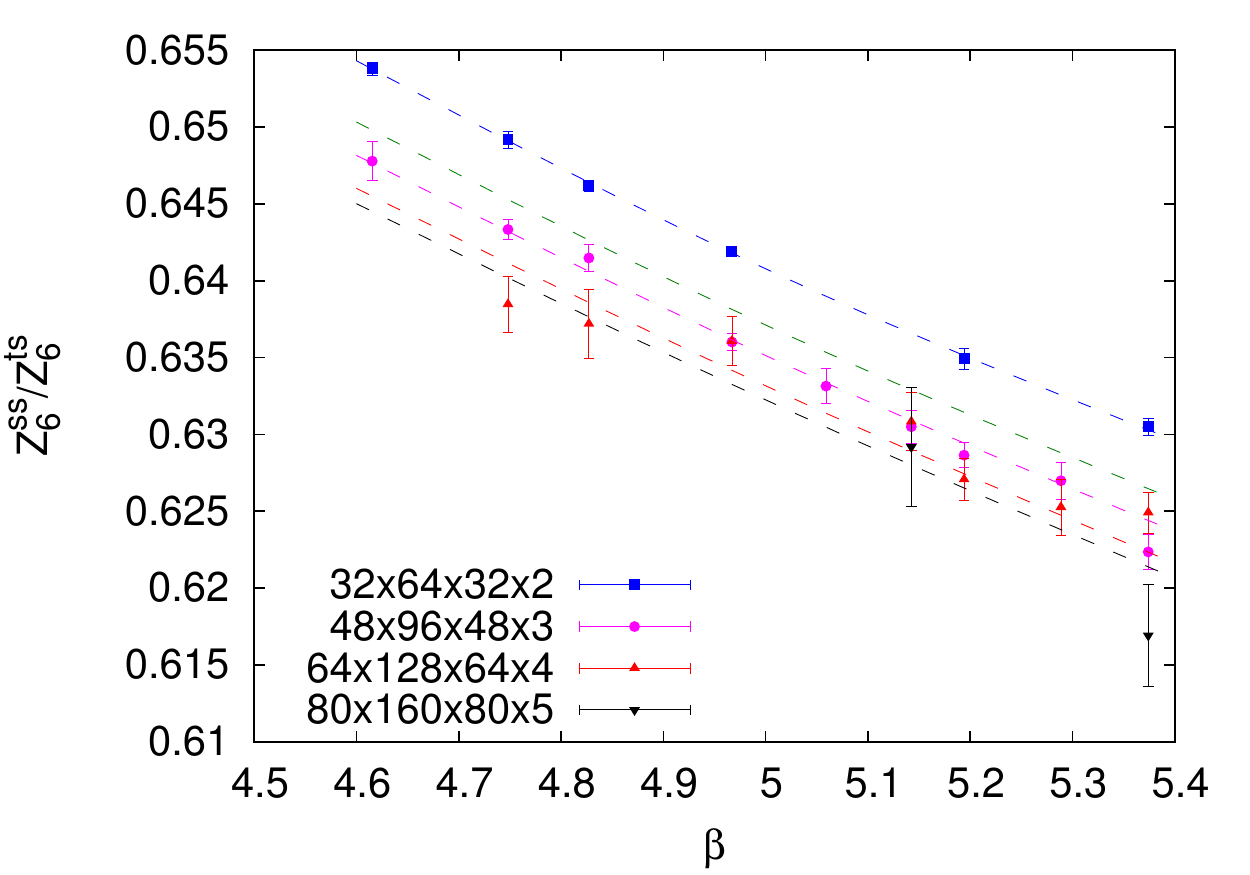} \\
\includegraphics[width=0.5\textwidth]{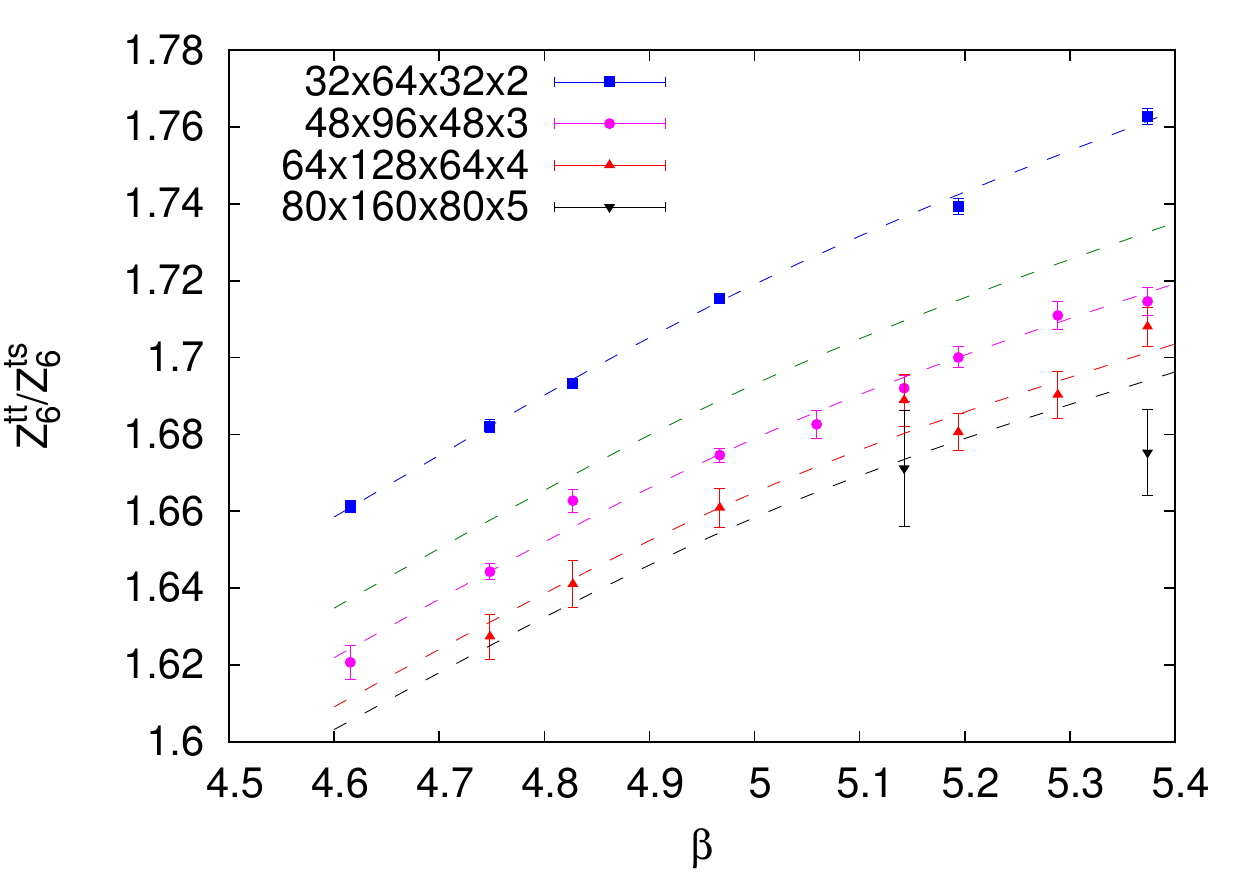}
\caption{
\label{fig:renorm}
The renormalization factors $Z^{ss}_{6}/Z^{ts}_{6}$ (top) and $Z^{tt}_{6}/Z^{ts}_{6}$ (bottom) as obtained from our
simulations with shifted boundary conditions with shift vector $\vec\xi=(1,1,1)$.}  
\end{center}
\end{figure}

For an isotropic gauge action the renormalization constants have been worked
out with shifted boundary conditions in \cite{Giusti:2015daa}. Using shifted
boundary conditions with shift vector $\vec\xi=(\xi_1,\xi_2,\xi_3)=(1,1,1)$ 
the off-diagonal $T_{0i}$ components develop a non-vanishing expectation value.
Since with this particular choice of the shift, the three spatial directions
are equivalent, we have $T_{01}=T_{02}=T_{03}$. Imposing this condition gives:
\begin{align}
2 Z^{tt}_6 \frac{1}{g_0^2}{F^a_{02}F^a_{12}} &= 2 Z^{ss}_6 \frac{1}{g_0^2}{F^a_{03}F^a_{13}} \\ \nonumber
                                             &= Z^{st}_6 \frac{1}{g_0^2}({F^a_{01}F^a_{21}}+{F^a_{03}F^a_{23}}) \rm{.}
\end{align}
Therefore, the ratios $Z^{ss}_{6}/Z^{ts}_{6}$ and $Z^{tt}_{6}/Z^{ts}_{6}$
can be calculated from a single simulation with 
$L_0^{-1}=T \sqrt{1+|\vec\xi|^2}=2T$.
Thus, e.g. to renormalize $T_{\mu\nu}$ in a $N_\tau = 12$ simulation with $\xi_R=2$,
we make an auxiliary run on a $48\times 96\times 48\times 3$ lattice with
the same bare parameters. The resulting factors will depend on
$\beta$ and $N_\tau$. The method requires that $N_\tau/4$ is an integer.
We observe a $1/N_\tau^2$ scaling of both $Z^{ss}_{6}/Z^{ts}_{6}$ and
$Z^{tt}_{6}/Z^{ts}_{6}$.  For the renormalization of $N_\tau=10$ we can
therefore use an interpolation in $N_\tau$. Our simulated results on the
renormalization factors can be seen in Figure~\ref{fig:renorm}.

The overall constant $Z^{ts}_6$ can be determined from the
following thermodynamic identity \footnote{See Appendix A}:
\begin{align}
\label{eq:srenorm}
C_{0101}(\tau,\mathbf{q=0})/T^5=- s/T^3 \rm{.}
\end{align}
This can be used for renormalization by requiring that the value of $C_{0101}$
at $\tau T = 0.5$ equals the continuum value of the entropy determined
in~\cite{Borsanyi:2012ve}. We used the values $s/T^3=5.02$ and $5.57$ for
$1.5T_c$ and $2T_c$ respectively.  This identity also provides a way to estimate
the order of magnitude of the discretization errors in $C_{0101}$. Since in the
continuum this correlator is independent of $\tau$, the $\tau$
dependence of the correlator gives a very direct way to see discretization
errors already on the finite $N_\tau$ data (for the case of $N_t=16$ 
see Fig.~\ref{fig:shear01}).

%\newpage

\section{Results on the correlators}
\subsection{Results at finite $N_t$}

\begin{figure}[h!]
\begin{center}
\includegraphics[width=0.5\textwidth]{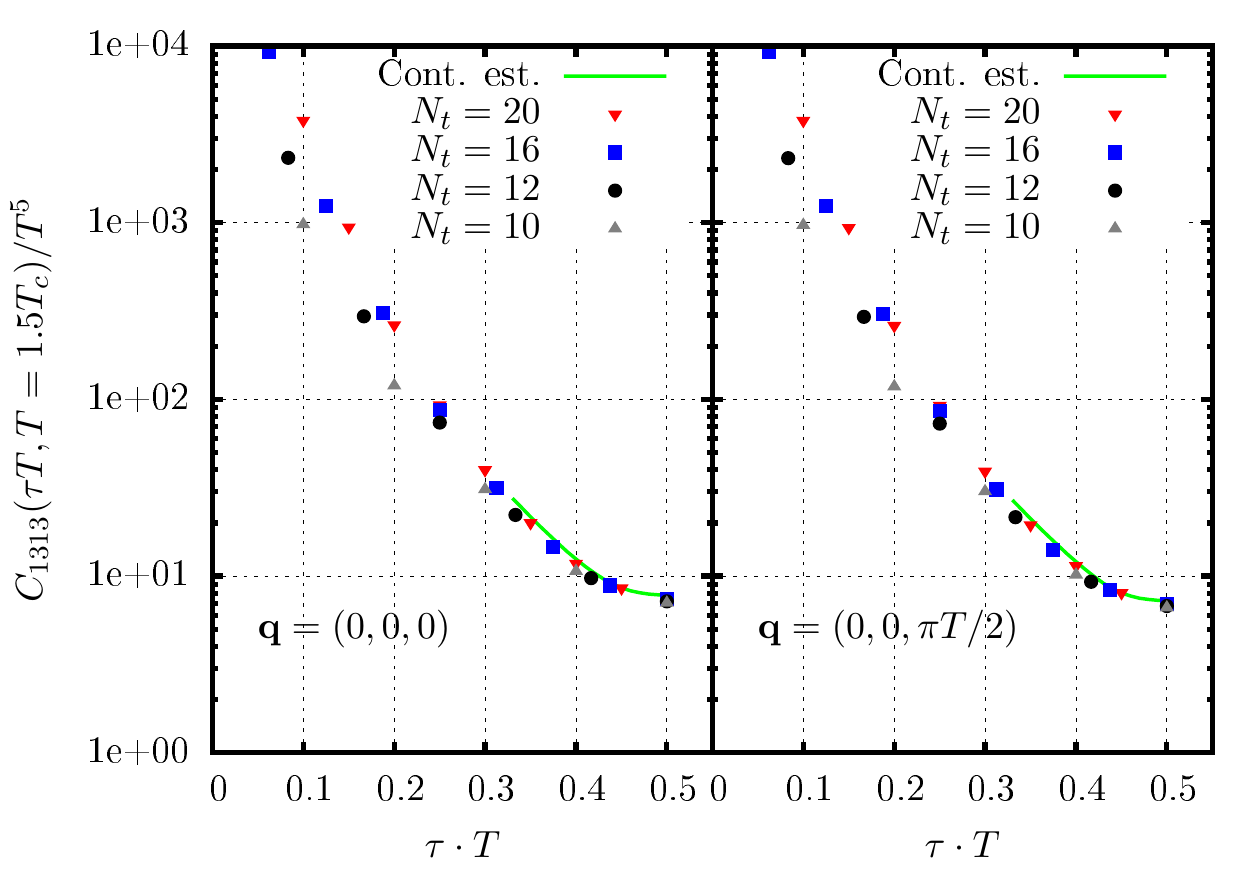} \\
\caption{
\label{fig:shear13} The renormalized shear correlator $C_{1313}$ at
different lattice spacings and different spatial momenta. We
also present a continuum estimate, that was produced by performing a spline
interpolation of the finite $N_t$ data. We only present the continuum estimate
in the range where the $\chi^2$ was acceptable.
}
\end{center}
\end{figure}

\begin{figure}[h!]
\begin{center}
\includegraphics[width=0.5\textwidth]{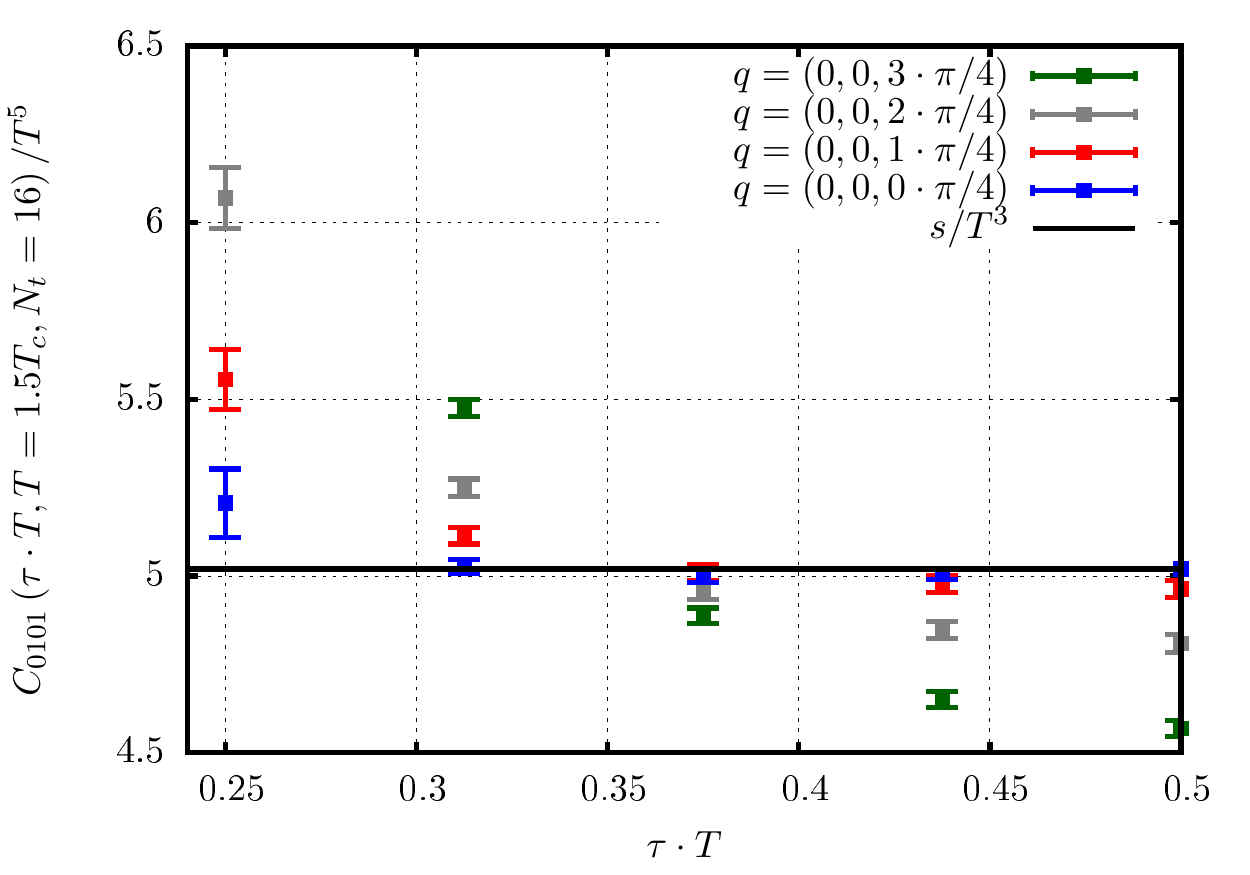} \\
\caption{
\label{fig:shear01} The renormalized shear correlator $C_{0101}$ at $N_t=16$ and for different spatial momenta for the temperature $T=1.5T_c$.
}
\end{center}
\end{figure}

The $13$ channel correlators can be seen in Fig.~\ref{fig:shear13}, while the 
results for the $01$ channel can be seen in Fig.~\ref{fig:shear01}.
For the $01$ channel, in the continuum, the correlator for zero spatial momentum 
should be a constant, equal to the entropy. The renormalization
condition we used for this correlator is simply that at the middle 
point, $\tau T = 1/2$ it should equal the continuum value of $-s/T^3$. 
How different the correlators value is for $\tau T \neq 1/2$ is some kind 
of measure of the cut-off effects. As we already discussed, we expect 
the $01$ channel to have smaller cut-off errors and also to be 
more sensitive to transport, so this is the more important of 
the two correlators. 

\subsection{Continuum limit extrapolation}

For the purpose of this paper we focus our discussion of the continuum limit
extrapolation to the middle point of the correlators $\tau T=1/2$.
We choose this approach for several reasons:
\begin{itemize}
  \item This is the most IR sensitive part of the correlators, therefore the most interesting part for studying transport.
  \item This is the part of the correlator with the least amount of cut-off effects, therefore one has to control the continuum extrapolation of this first,
 before attempting to go to smaller separations in imaginary time.
\end{itemize}
Notice, that the $C_{1313}$ correlator is closely related to the $\tau$ derivative of the $C_{0101}$ correlator: %. This can be seen if one looks at formulas:
\begin{align}
\frac{d^2 C_{0101}(\tau =1/2T, \mathbf{q})}{d\tau^2} &= \int d \omega  \frac{\omega^2 \rho_{0101}(\omega,\mathbf{q})}{\sinh \left( \frac{\beta \omega}{2} \right)} \\ 
C_{1313}(\tau =1/2T, \mathbf{q})                     &= \int d \omega  \frac{-(\omega^2 / \mathbf{k}^2) \rho_{0101}(\omega, \mathbf{q})}{\sinh \left( \frac{\beta \omega}{2} \right)} \\
C_{1313}(\tau =1/2T, \mathbf{q})                     &= -\frac{1}{\mathbf{k}^2} \frac{d^2 C_{0101}(\tau =1/2T, \mathbf{q})}{d\tau^2} \rm{,} 
\end{align}
as can be seen from a differentiation of the sum rule (\ref{eq:sumrule}) and
application of the Ward identity (\ref{eq:Ward}) respectively. 
Thus, taking the $C_{1313}(\tau,\mathbf{q})$ and $C_{0101}(\tau,\mathbf{q})$
correlators only at the value $\tau T=1/2$ already contains the
leading $\tau$ dependence of $C_{0101}$. Thus, we may continue with the
extrapolation at $\tau T=1/2$.

\begin{figure}[h!]
\begin{center}
\includegraphics[width=0.5\textwidth]{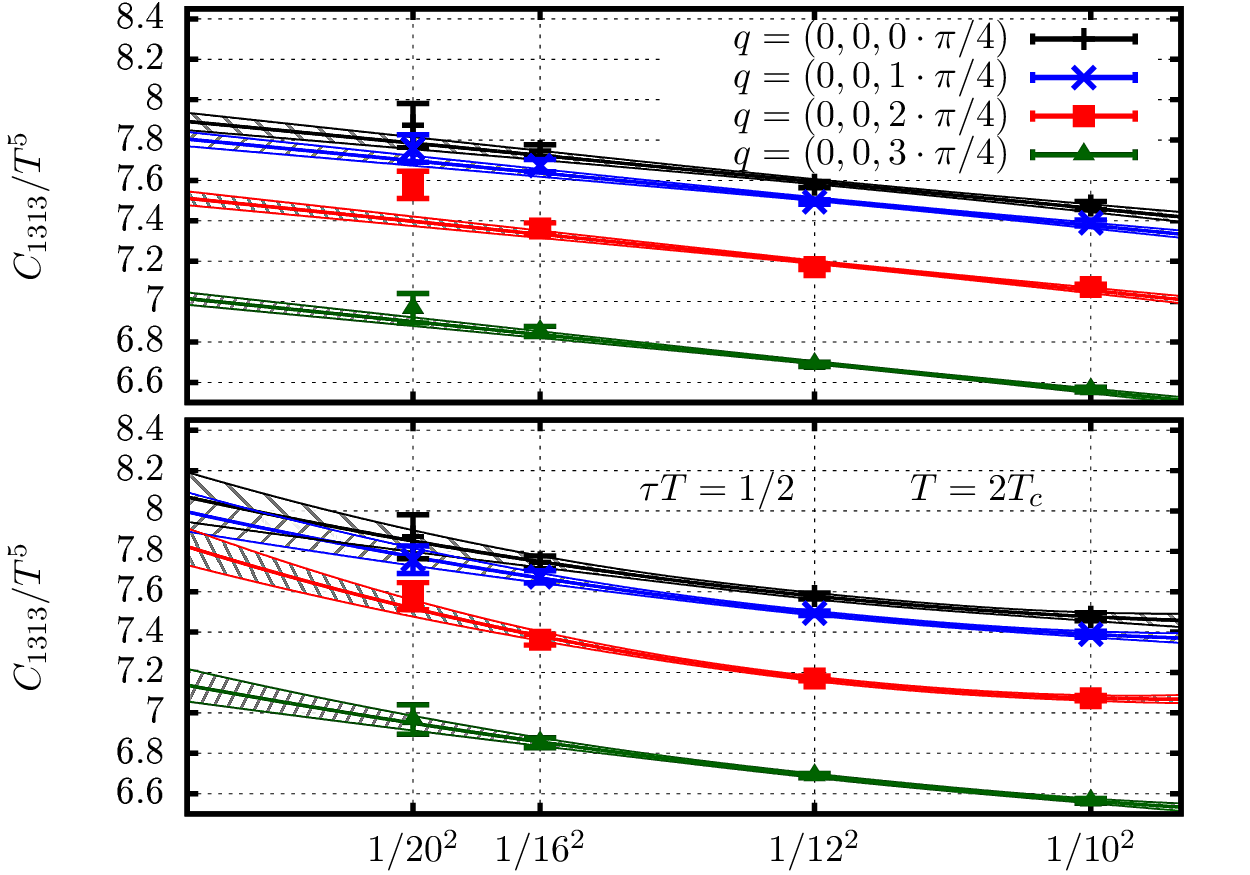} \\
\caption{
    \label{fig:clim13} Continuum limit extrapolation of $C_{1313}$ at $\tau T=1/2$ and $T=2T_c$. The tree level improvement was not applied to the data. 
    Top: linear fit; Bottom: quadratic fit
}
\end{center}
\end{figure}

\begin{figure}[h!]
\begin{center}
\includegraphics[width=0.5\textwidth]{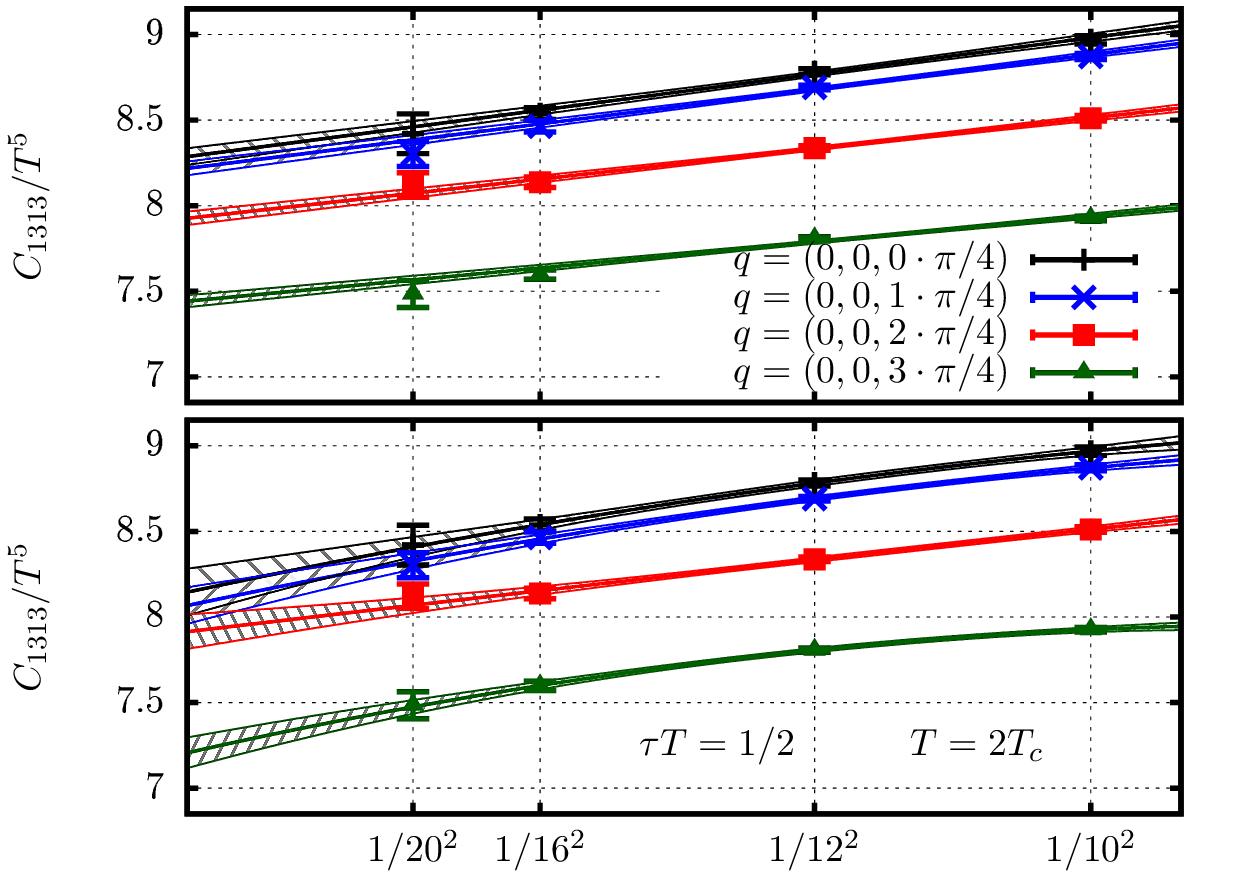} \\
\caption{
    \label{fig:clim13_tli} Continuum limit extrapolation of $C_{1313}$ at $\tau T=1/2$ and $T=2T_c$ with the tree level improvement applied to the lattice data. 
    Top: linear fit; Bottom: quadratic fit
}
\end{center}
\end{figure}

\begin{figure}[h!]
\begin{center}
\includegraphics[width=0.5\textwidth]{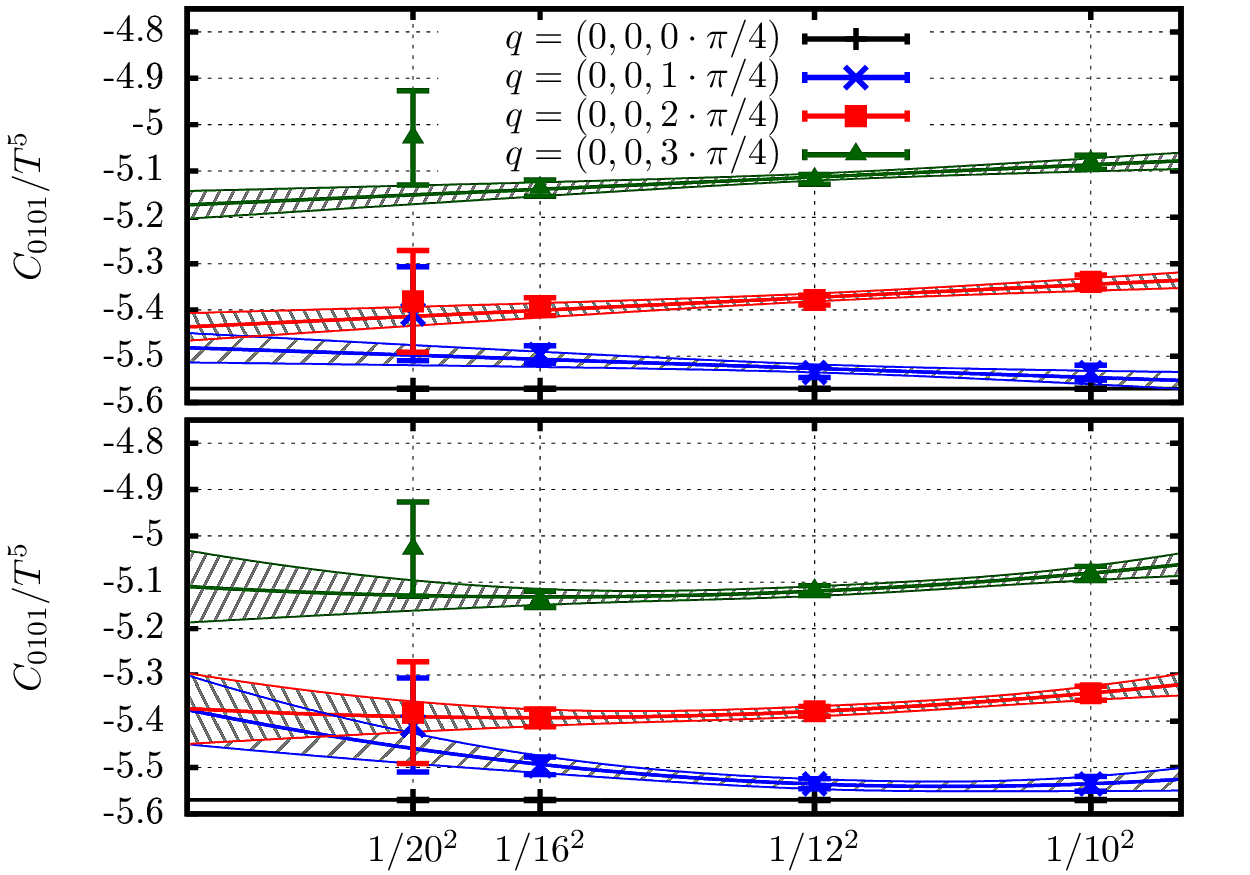} \\
\caption{
    \label{fig:clim01} Continuum limit extrapolation of $C_{0101}$ at $\tau T=1/2$ and $T=2T_c$. %The tree level improvement was not applied to the data. 
    Top: linear fit; Bottom: quadratic fit
}
\end{center}
\end{figure}

We will attempt a continuum limit extrapolation both with and without tree
level improvement. The tree level improvement coefficients are the result of a
tedious, but straightforward computation. The numerical values of the
improvement coefficients are summarized in Appendix B. We will also attempt
both linear and quadratic fits for the continuum limit extrapolation.
Attempting a continuum limit extrapolation from our $N_t=10,12,16,20$ data
yields the following behavior:
\begin{itemize}
\item In the $0101$ channel, since one applies the renormalization condition (11) after the tree level improvement, the continuum extrapolation
 is quite flat, regardless of whether one uses tree level improvement or not.
\item A linear fit to the $N_t=10,12,16,20$ lattices in the $1313$ channel with and without tree level improvement does 
  not always yield consistent results within 1$\sigma$ for the continuum limit extrapolation.
\item A quadratic fit to the $N_t=10,12,16,20$ lattices in the $1313$ channel with and without tree level improvement does yield consistent results, but 
then we have one degree of freedom less, so the error on the continuum
is larger, roughly on the $2-3\%$ level.
\item Linear versus quadratic fits to the data obtained without tree level improvement 
      are not consistent within $1\sigma$ for the $1313$ channel.
\item Linear versus
quadratic fits to the tree level improved data are closer, but still not
consistent within $1\sigma$ for the $1313$ channel.
\end{itemize}
This behavior can be visually observed in 
Figures \ref{fig:clim13}, \ref{fig:clim13_tli} and \ref{fig:clim01} where 
the linear and quadratic extrapolations are shown.

From this analysis, we conclude that from our present data, the continuum extrapolation has 
errorbars on the few percent level, for both channels. The results for the 3 point linear fits 
are summarized in Table \ref{table:corr_clim}.
\begin{table}
\begin{center}
\begin{tabular}{ | c | c | c | c |  }
\hline
channel & $q_3$ & result for $1.5T_c$ & result for $2T_c$\\
\hline
$01$    &  $\pi T/4$    & -4.93(7)  & -5.41(7) \\
$01$    &  $\pi T/2$    & -4.66(10) & -5.40(6) \\
$01$    &  $3 \pi T/4$  & -4.55(9)  & -5.15(5) \\
\hline
$13$    &  $0$          & 7.83(13)  & 8.12(15) \\
$13$    &  $\pi T/4$    & 7.47(8)   & 8.04(13) \\
$13$    &  $\pi T/2$    & 7.24(10)  & 7.90(7)  \\
$13$    &  $3 \pi T/4$  & 6.70(7)   & 7.15(11) \\
\hline
\end{tabular}
\caption{\label{table:corr_clim} Values of the correlators in the continuum. The error bar includes statistical
    errors, as well as systematic errors coming from the linear vs quadratic continuum fit, and continuum extrapolation
    with and without tree level improvement.
}
\end{center}
\end{table}

\subsection{Finite volume effects at tree level}

From the tree level calculation we can estimate the finite volume effects on the UV contribution to the correlators. 
For the volumes used for
our simulations, i.e. $L_xT=L_yT=2$ and $L_zT=8$ we 
calculated the tree-level (UV) contribution of the spectral
function in Appendix~B. The relative deviation from the infinite volume
contribution is shown in Table~\ref{table:finiteV}.
Thus, the tree level finite volume effect on the observables considered
here is on the $10\%$ level. While $10\%$ error on the final viscosity
is probably harmless at this point, it may shift the relative
weight of the UV and IR contributions. 
It is important to note, that the finite volume correction
depends very weakly on $q$ and for $C_{0101}$ it weakly depends on $\tau$, too.
Thus, in the viscosity fits the volume dependence approximately factorizes, and  it
affects only the value of the $c$ parameter
in Eq.~(\ref{eq:Cfit}). The values we quote later will correspond
to the raw fit, which is expected to be roughly 10\% below the infinite volume
value.

\begin{table}
\begin{center}
\begin{tabular}{ | c | c | c | c |  }
\hline
channel & $\tau T$ &  $q_3$       & (finite vol.)/(infinite vol.) \\
\hline
$01$    &   $1/2$  &  $0$         & 0.90 \\
$01$    &   $1/2$  &  $\pi T/4$   & 0.92 \\
$01$    &   $1/2$  &  $\pi T/2$   & 0.90 \\
$01$    &   $1/2$  &  $3 \pi T/4$ & 0.89 \\
\hline
$13$    &   $1/2$  &  $0$         & 0.89 \\
$13$    &   $1/2$  &  $\pi T/4$   & 0.90 \\
$13$    &   $1/2$  &  $\pi T/2$   & 0.90 \\
$13$    &   $1/2$  &  $3 \pi T/4$ & 0.90 \\
\hline
$01$    &   $1/4$  &  $0$         & 0.90 \\
$01$    &   $1/4$  &  $\pi T/4$   & 0.92 \\
$01$    &   $1/4$  &  $\pi T/2$   & 0.91 \\
$01$    &   $1/4$  &  $3 \pi T/4$ & 0.90 \\
\hline
$13$    &   $1/4$  &  $0$         & 0.99 \\
$13$    &   $1/4$  &  $\pi T/4$   & 0.99 \\
$13$    &   $1/4$  &  $\pi T/2$   & 0.99 \\
$13$    &   $1/4$  &  $3 \pi T/4$ & 0.99 \\
\hline
\end{tabular}
\caption{\label{table:finiteV}Finite volume corrections at tree level for the different correlators.
}
\end{center}
\end{table}

\section{Estimating the viscosity}
%The present study, like studies before it could not achieve the desired high accuracy needed to 
%differentiate between different scenarios (featureless vs quasi particle peak) for the spectral 
%function. For this reason, 
To get an educated guess on the viscosity, one needs to assume an ansatz.
Here, we assume a very simple hydrodynamics plus tree level ansatz for the
spectral function, corresponding to the featureless scenario in Figure 1:
\begin{align}
\label{eq:Cfit}
C_{\mu \nu \mu \nu}(\tau,\mathbf{q}) &= c \int_{|\mathbf{q}|}^\infty \rho_{\mu \nu \mu \nu}^{\rm{tree\ level}}(\omega,\mathbf{q}) K(\tau,\omega) d\omega \\ \nonumber 
 &+ \int_0^\infty \rho_{\mu \nu \mu \nu}^{\rm{hydro}}(\omega,\mathbf{q};\eta/s) K(\tau, \omega) d \omega
\end{align}
The hydrodynamic predictions for the spectral function are written out in
Appendix A. The tree level spectral function is taken to be the one at infinite
volume and in the continuum. Notice that the integral in $\omega$ for this UV
part is cut off at $\omega=|\mathbf{q}|$ in the IR. 
The part at lower $\omega$ is responsible for the hydrodynamical
behaviour, which is taken into account by the other term. Actually,
the free gas formula would give an infinite contribution to the viscosity.
The formulas for the continuum tree level spectral function can be found in
Ref.~\cite{Meyer:2008gt} and are also summarized in Appendix B.

We introduce a constant $c$ in front of the spectral
function. 
We do so to account in a simple way for higher order 
and also finite volume corrections. The assumption that all of these effects
can be put into a single constant is a very strong one. One hint that it might
be a good estimate is given by Table II, where we have the finite volume
correction factors for different correlators.

This model has two free parameters
\footnote{The entropy is fixed from previous calculations of the equation of state.}, 
the factor $c$ in front of the tree level correlator, and the shear viscosity to
entropy ratio $\eta/s$ may be contaminated by higher order contributions, as
well. Clearly, our estimate of the viscosity is correct only in the
range of validity of this simple model for the spectral function.
In the following, we will work out how data constrain the model parameters.
Assuming that our system is within the model's range of validity we can
make a quantitative statement on the shear viscosity.

\subsection{Sensitivity to model parameters}
\begin{figure}[th!]
\begin{center}
\includegraphics[width=0.5\textwidth]{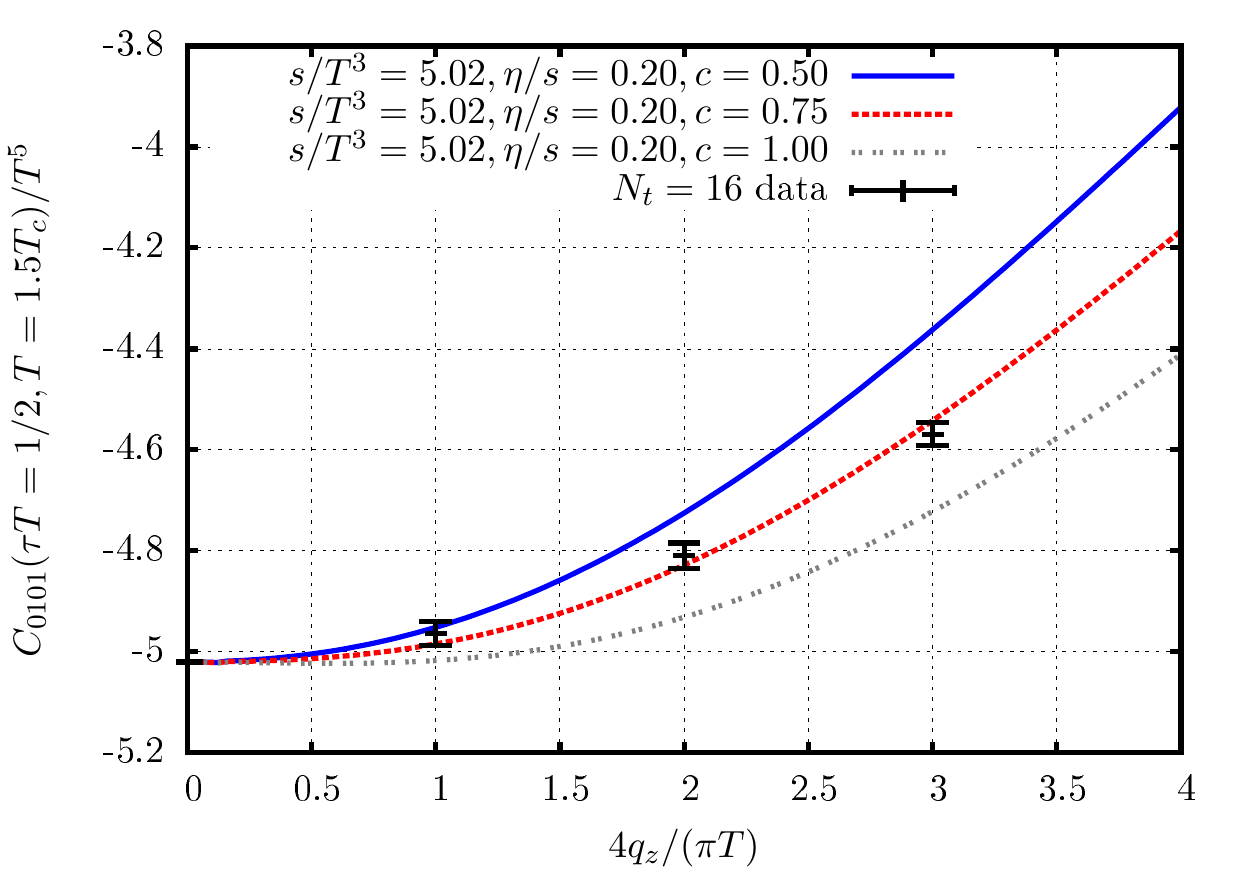} \\
\caption{
    \label{fig:changeC} Effect of changing the UV parameter $c$ in the model on the correlator $C_{0101}$ at $\tau T=1/2$ for several different
    spatial momenta.
}
\end{center}
\end{figure}

\begin{figure}[th!]
\begin{center}
\includegraphics[width=0.5\textwidth]{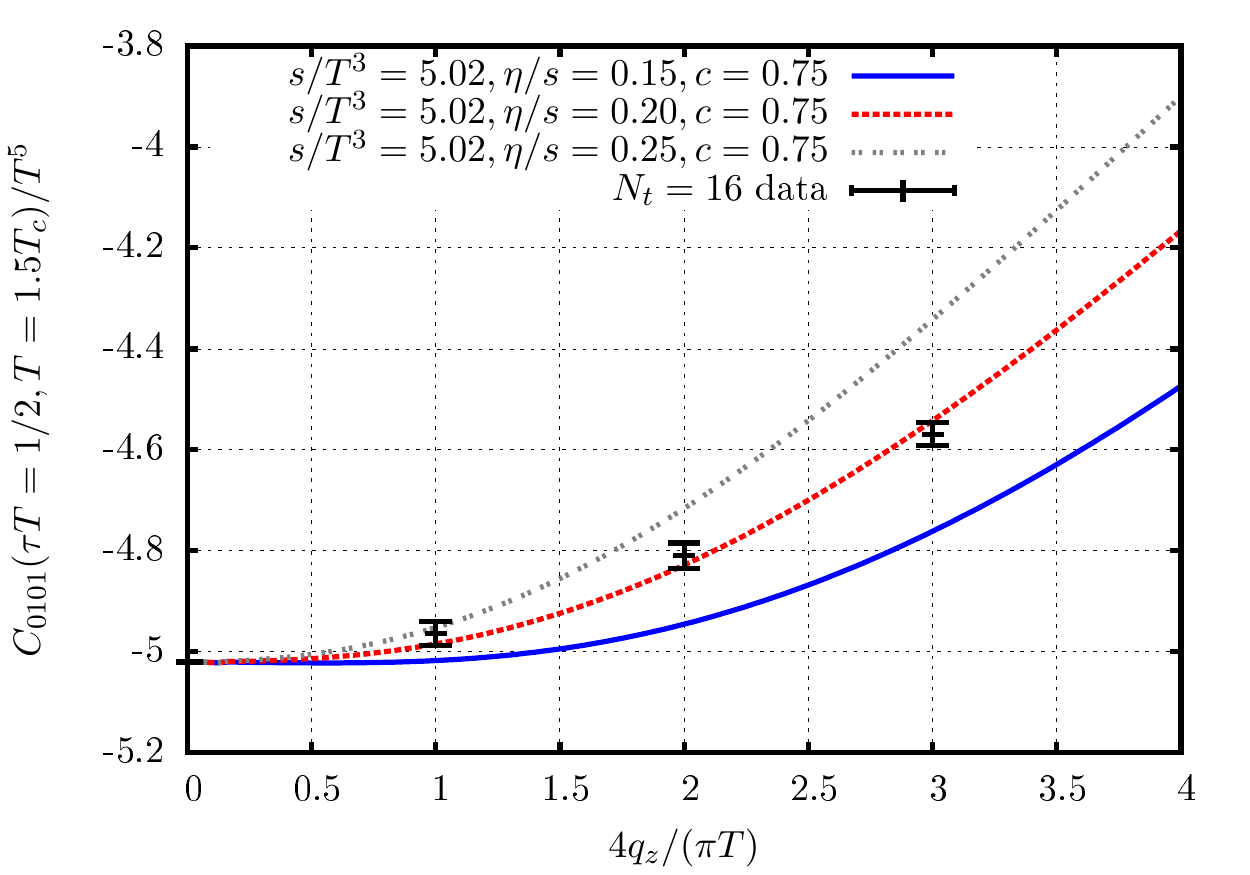} \\
\caption{
    \label{fig:changeetas} Effect of changing the hydrodynamic parameter $\eta/s$ in the model on the correlator $C_{0101}$ at $\tau T=1/2$ for several different spatial momenta.}
\end{center}
\end{figure}

Before describing the fitting procedure let us show how the
different model parameters influence the observables considered here.  Since it
is the more interesting quantity, our discussion here will focus on $C_{0101}$.
Figures \ref{fig:changeC} and \ref{fig:changeetas} concentrate on changing one
of the parameters, $c$ or $\eta/s$, respectively. From these pictures the
conclusion one can draw is that $C_{0101}$, while certainly sensitive to the
hydrodynamic parameter $\eta/s$, is also sensitive to the UV parameter $c$.
This is not surprising, but it is a big advantage compared to the $C_{1313}$
correlator, where the sensitivity to $\eta/s$ is smaller. Still, one has to
acknowledge that while this is a pretty useful quantity, because of the 
sensitivity to both parameters, it is not enough to
constrain the value of $c$ and $\eta/s$ together. To do that one has to
consider in addition the $\tau$ dependence of $C_{0101}$, or equivalently, the
correlator $C_{1313}$ at $\tau=0$, which, as we have already shown, corresponds to
the second $\tau$ derivative of $C_{0101}$.

\subsection{Fits at finite $N_t$}
We will present two different fits for the parameter $c$ and $\eta/s$. First we study only $C_{0101}$ as a function of $\tau$ and $\mathbf{q}$ 
at $N_\tau=16$. 
This approach is similar to what was used in earlier publications, 
where only data at one finite $N_t$ were available. 
Our choice of the channel $C_{0101}$ is motivated by the smaller cut-off errors compared to $C_{1313}$, as well as its higher
sensitivity to the transport part of the spectral function. We constrain our
fits to the range $\tau T \in [0.3,0.5]$. This range
is motivated by two observations: 
i) the quantity $C_{0101}(\tau,\mathbf{q}=0)$, which is, in principle, 
independent of $\tau$ in the continuum, 
is indeed constant for $N_t=16$ in this range (see Fig.~\ref{fig:shear01});
ii) the finite volume corrections at tree level are also $\tau$ independent 
in this range.
The latter fact motivates the assumption that most finite volume effects
can be captured by a modified value of the $c$ parameter.

\begin{center}
  \begin{tabular}{c | c | c }
    $T$      & $\eta/s$     & $c$         \\
    \hline
    $1.5T_c$ & $0.178(15)$  & $0.60(6)$   \\
    $2.0T_c$ & $0.157(13)$  & $0.63(7)$   \\
  \end{tabular}
\end{center}
Here the error includes statistical errors, as well as systematic
errors coming from the choice of $\tau_{min}=5./16$ or $6./16$ and 
$q_{max}=2\pi/4$ or $q_{max}=3\pi/4$.
These numbers are, of course, only valid once we assume our 
hydrodynamic ansatz.

The viscosity appears to be temperature independent from our analysis. Here we have to mention
a serious drawback of our fit ansatz. It assumes that the hydrodynamic
prediction for the spectral function, strictly valid only for $\omega \ll T$,
is also a good approximation for higher frequencies. This is true for $\mathcal{N}=4$ SYM
theory, where AdS/CFT can be used to calculate the spectral function
\cite{Teaney:2006nc}. Our ansatz cannot produce a quasiparticle peak, that
would appear in weak coupling treatements of QCD, like kinetic theory
\cite{Arnold:2000dr,Arnold:2003zc,Hong:2010at}. This means that the physical
mechanism that makes the viscosity diverge for $T \to \infty$, namely the
sharpening of the peak in $\rho_{1313}$ near $\omega=0$ is missing from our
ansatz. This implies that our ansatz can certainly not be used at very high
temperatures, where the weak coupling calculation is trustworthy, and even at
intermediate temperatures we might underestimate the viscosity, effectively
smearing out the transport peak by enforcing the ansatz in the data analysis.  
This is a weakness shared by all previous lattice estimates of the shear
viscosity, since they either use a very similar hydrodynamic ansatz, or the 
Backus-Gilbert method, in which case the base functions used for the reconstruction 
similarly prefer a featureless behavior, as opposed to a transport 
peak~\cite{Meyer:2011gj}. 

\subsection{Fits on continuum data}
For the second fit we consider the $q_3/(\pi T/4)=0,1,2,3$ dependence of 
$C_{0101}(\tau T=0.5)$ and $C_{1313}(\tau T=0.5)$ in the continnum. Our
results are:

\begin{center}
\begin{tabular}{c | c | c }
$T$      & $\eta/s$  & $c$         \\
\hline
$1.5T_c$ & $0.17(2)$ & $0.63(3)$   \\
$2.0T_c$ & $0.15(2)$ & $0.67(3)$   \\
\end{tabular}
\end{center}
Here, the error is statistical only. The systematic error 
coming from the choice of the $\tau$ range does not exist here, since we always
use just $\tau=1/2$. 
This fit uses the same hydrodynamic ansatz as discussed above.

This is the first estimate of $\eta/s$ using continuum extrapolated data. It is
consistent with earlier estimates using a single lattice spacing
\cite{Meyer:2007ic,Mages:2015rea}.

\section{Summary}
In this paper we studied the continuum behavior of the energy-momentum tensor
correlators in pure $SU(3)$ gauge theory. We found cut-off errors at a few percent
level for $N_t=16$. For some quantities, namely $C_{1313}$ and $C_{0101}$ at
several spatial momenta, and $\tau T = 1/2$ continuum extrapolation was
possible. Out of these quantities $C_{0101}$ is actually sensitive to the
transport part of the spectral function.  

The achieved percent level precision of the data does not yet allow us to
distinguish different scenarios for the spectral function. The precision
was already boosted by the multi-level algorithm on an anisotropic lattice.
Despite the promising extension of the multi-level algorithm to full QCD
\cite{Ce:2016idq} it is hardly possible to achieve even this precision  
with dynamical quarks. Thus, for the testing of the hydrodynamical model
one has to seek for alternative methods.

Once, however, a model is postulated, it is possible to give a model dependent
estimate of the shear viscosity to entropy ratio. We gave the first estimate of
this phenomenologically important quantity from continuum extrapolated lattice
data.  Our estimate is in the same ballpark as earlier estimates based on
finite $N_t$ lattices.

\section*{Acknowledgments}
This project was funded by the DFG grant SFB/TR55. 
This work was partially supported by the Hungarian National Research,
Development and Innovation Office –- NKFIH grants KKP126769 and K113034.  
An award of computer time was provided by the INCITE
program. This research used resources of the Argonne Leadership Computing
Facility, which is a DOE Office of Science User Facility supported under
Contract DE-AC02-06CH11357. We acknowledge computer time on the QPACE machines
\cite{Baier:2009yq} on the Wuppertal and Jülich sites. The authors gratefully
acknowledge the Gauss Centre for Supercomputing (GCS) for providing computing
time for a GCS Large-Scale Project on the GCS share of the supercomputer
JUQUEEN at the Jülich Supercomputing Centre \cite{juqueen}.

This material is based upon work supported by the National Science Foundation 
under grants no. PHY-1654219 and OAC-1531814 and by the U.S. Department of Energy, 
Office of Science, Office of  Nuclear  Physics,  within  the  framework  of  the  
Beam Energy Scan Theory (BEST) Topical Collaboration. C.R. also acknowledges the 
support from the Center of Advanced Computing and Data Systems at the 
University of Houston.

\section*{Appendix A: Predictions of hydrodynamics for the spectral functions}

The combination of linearized relativistic hydrodynamics and linear response theory allows 
for a derivation of the low frequency behavior of the energy-momentum tensor correlators.
For a nice derivation of these formulas, see the Appendix of~\cite{Teaney:2006nc}. Here, we just
collect the relevant formulas in our notation, for easy reference. We assume the spatial momentum is
in the $z$ direction $\mathbf{k} = (0,0,k)$. In this case the spectral functions in the shear channel are:
\begin{align}
\label{eq:hydro1}
-\frac{\rho_{0101}}{\omega} = \frac{\eta}{\pi} \frac{k^2}{ \omega^2 + \left(\frac{\eta}{sT} k^2\right)^2} \\
\label{eq:hydro2}
\frac{\rho_{1313}}{\omega} = \frac{\eta}{\pi} \frac{\omega^2}{ \omega^2 + \left(\frac{\eta}{sT} k^2\right)^2} \rm{,}
\end{align}
where $s$ is the entropy, $\eta$ is the shear viscosity and $T$ is the temperature. We use these formulas for both our finite $N_\tau$
and continuum data fits. The zero spatial momentum limit of $\rho_{1313}/\omega$ is a constant equal to $\eta/\pi$, 
while the zero spatial momentum limit of the $\rho_{0101}/\omega$ is a delta function at the origin:
\begin{align}
 \frac{\rho_{1313}}{\omega} \to \frac{1}{2} s T \delta(\omega - \epsilon) \rm{,}
\end{align}
as can be easily shown using equation (\ref{eq:hydro1}). This is the hydrodynamic identity we utilize for our
renormalization procedure. Formulas (\ref{eq:hydro1}) and (\ref{eq:hydro2}) are also the basis for the derivation of the 
Kubo formulas, like equation (1).

\section*{Appendix B: Tree level spectral function in the continuum}
The leading order perturbative result for the spectral function at high frequency is~\cite{Meyer:2008gt}:
\begin{align}
-\rho_{0101}^{(pert)}=&\frac{d_A}{8 (4 \pi)^2} q^2 (\omega^2 - q^2) \mathcal{I}(\omega,q,T) \rm{,} \\
\mathcal{I}(\omega,q,T) =& \theta(\omega-q)\int_0^1 dz \frac{(1-z^4) \sinh(\omega/2T)}{\cosh(\omega/2T)-\cosh(qz/2T)} +  \nonumber \\
+& \theta(-\omega+q)\int_1^\infty  \frac{(z^4-1) \sinh(\omega/2T)}{\cosh(\omega/2T)-\cosh(qz/2T)} \rm{.}
\end{align}
The tree level result in the $\rho_{1313}^{pert}$ channel follows trivially
using the Ward identity (\ref{eq:Ward}).

In our analysis, we only take the first part $\omega > q$. The reason for that is that the $\omega<q$ part describes the transport properties of 
a free gas of gluons, and corresponds to an infinite viscosity. 
We therefore drop this term and substitute it with the ansatz from hydrodynamics 
(which describes a strongly coupled system). 

\section*{Appendix C: Tree level improvement coefficients}

The formulas for the tree level improvement are the result of a tedious but
straightforward leading order calculation. The resulting formulas still
contain Matsubara sums, that can be easily evaluated numerically. For
reference, we include here the numerical values of the tree level improvement
coefficients relevant for our study.

\hspace{10cm}

\begin{center}
\begin{tabular}{ | c | c | c | c | }
\hline
    $mn$ & $q \frac{4}{\pi T}$ & $N_t$ & $C^{tl}(N_t=\inf)/C^{tl}(N_t)$ \\
    \hline
    01 & 0 &  10  & 1.26852 \\
    01 & 0 &  12  & 1.19188 \\
    01 & 0 &  16  & 1.11254 \\
    01 & 0 &  20  & 1.07385 \\
\hline
    01 & 1 &  10  & 1.25721 \\
    01 & 1 &  12  & 1.18425 \\
    01 & 1 &  16  & 1.10833 \\
    01 & 1 &  20  & 1.07118 \\
\hline
    01 & 2 &  10  & 1.26208 \\
    01 & 2 &  12  & 1.18749 \\
    01 & 2 &  16  & 1.11007 \\
    01 & 2 &  20  & 1.07227 \\
\hline
    01 & 3 &  10  & 1.27067 \\
    01 & 3 &  12  & 1.19333 \\
    01 & 3 &  16  & 1.11328 \\
    01 & 3 &  20  & 1.07430 \\
\hline
    13 & 0 &  10  & 1.19957 \\
    13 & 0 &  12  & 1.15874 \\
    13 & 0 &  16  & 1.10223 \\
    13 & 0 &  20  & 1.06957 \\
\hline
    13 & 1 &  10  & 1.20054 \\
    13 & 1 &  12  & 1.15971 \\
    13 & 1 &  16  & 1.10297 \\
    13 & 1 &  20  & 1.07011 \\
\hline
    13 & 2 &  10  & 1.20329 \\
    13 & 2 &  12  & 1.16254 \\
    13 & 2 &  16  & 1.10515 \\
    13 & 2 &  20  & 1.07168 \\
\hline
    13 & 3 &  10  & 1.20742 \\
    13 & 3 &  12  & 1.16703 \\
    13 & 3 &  16  & 1.10869 \\
    13 & 3 &  20  & 1.07428 \\
%    13 & 4 &  10  & 1.21226 \\
%    13 & 4 &  12  & 1.17281 \\
%    13 & 4 &  16  & 1.11348 \\
%    13 & 4 &  20  & 1.07783 \\
\hline
\end{tabular}
\end{center}

\newpage

\bibliography{visc}
\bibliographystyle{apsrev4-1}

\end{document}